\documentclass[
preprint, 
superscriptaddress,
 amsmath,amssymb,
 aps, 
floatfix,
pre,
]{revtex4-1}

\usepackage{graphicx}
\usepackage{dcolumn}
\usepackage{bm}
\usepackage{pythonhighlight}
\usepackage{xcolor}

\begin{document}

\title{Controlling sternness in judging a good person who helps the bad}%

\author{Quang Anh Le}
\thanks{These two authors contributed equally.}
\affiliation{Industry-University Cooperation Foundation, Pukyong National
University, Busan 48513, South Korea}
\author{Minwoo Bae}
\thanks{These two authors contributed equally.}
\affiliation{Department of Complexity Science and Engineering, The University of Tokyo, Chiba 277-8561, Japan}
\author{Takashi Shimada}
\email[]{shimada@sys.t.u-tokyo.ac.jp}
\affiliation{Department of Systems Innovation, Graduate School of Engineering, The University of Tokyo, 7-3-1 Hongo, Bunkyo-ku, Tokyo 113-8656, Japan}
\author{Seung Ki Baek}
\email{seungki@pknu.ac.kr}
\affiliation{Department of Scientific Computing, Pukyong National University,
Busan 48513, South Korea}
\date{\today}

\begin{abstract}
Recent studies on indirect reciprocity with private assessment on complete graphs suggest the possibility that one can continuously modulate the degree of segregation by controlling how to judge a good person helping a bad one. A well-known social norm called L6 judges it as bad, which eventually segregates the society into two antagonistic clusters, but if it is judged as good, the system reaches paradise where everyone likes each other.
In this work, we numerically study this transition between segregation and paradise in two different settings. Firstly, in a uniform population of size $N$ where everyone regards such a donor as good with probability $p$ and bad with $1-p$, we observe paradise when $Np$ is sufficiently greater than $O(1)$. In contrast, in a heterogeneous setting where only $k$ individuals judge such a donor as good, the size difference of the clusters increases almost linearly as $k$ increases, so paradise can only be reached as $k \to N$ in a large population. Therefore, when an urgent change is needed to overcome the segregation due to L6, a small change in each and every individual's behavior is more efficient than a radical change in a fraction of the population.
\end{abstract}

\maketitle


\section{Introduction}

\begin{figure}
\includegraphics[width=0.45\columnwidth]{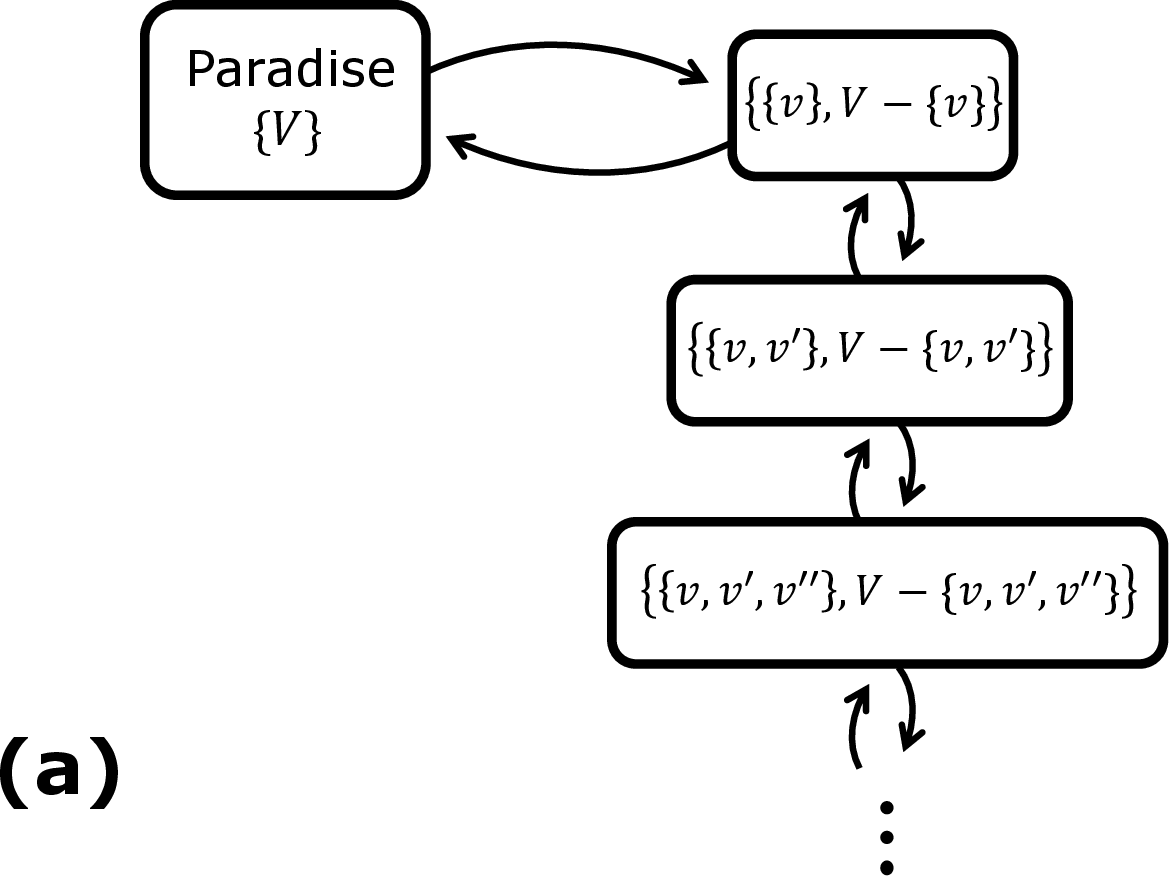}
\includegraphics[width=0.45\columnwidth]{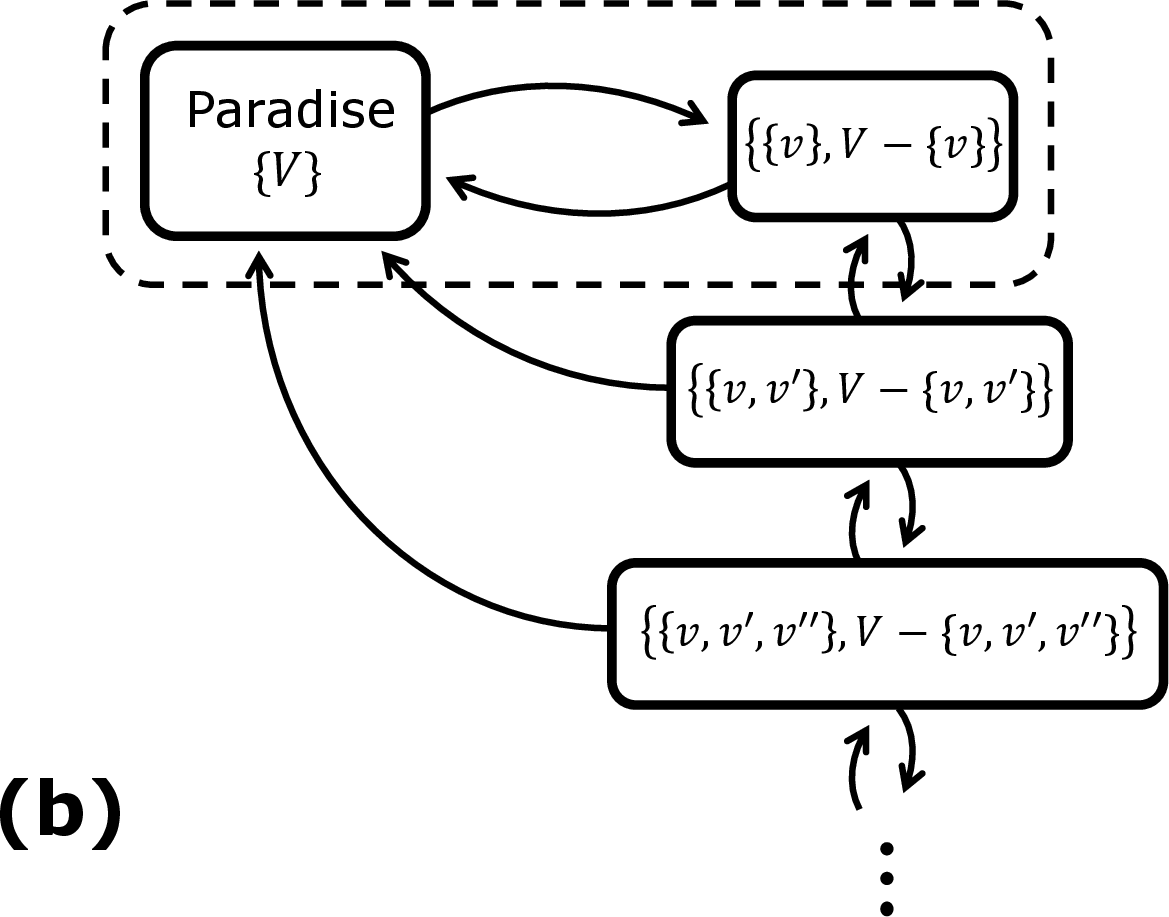}
\caption{Transitions induced by (a) L6 and (b) L4 among balanced configurations consisting of two clusters. The dashed box in (b) shows an outflow from paradise, but it becomes vanishingly weak compared to the inflow to paradise as $N$ increases~\cite{bae2024exact}.}
\label{fig:transition}
\end{figure}

Recent studies on indirect reciprocity~\cite{oishi2013group,oishi2021balanced,bae2024exact} have revealed its close connection to structural balance theory~\cite{heider1946attitudes,harary1953notion,cartwright1956structural}.
More specifically, L6 (Stern Judging)~\cite{pacheco2006stern} has been shown to induce a balanced configuration with two antagonistic clusters of almost equal sizes in a complete graph.
They have opinions about each other, and an individual $i$'s opinion about $j$ is indicated by $\sigma_{ij}$. If $i$ thinks of $j$ as good, $\sigma_{ij}=+1$, and $-1$ otherwise.
In the donation game between a donor and a recipient, which are indicated by $d$ and $r$, respectively, if $d$ regards $r$ as good, that is, if $\sigma_{dr}=+1$, $d$ helps $r$ at a cost of $c$, resulting in a benefit of $b$ on the recipient's side. Otherwise, $d$ does nothing to $r$. The interaction between $d$ and $r$ is observed by everyone in the population, who judges the donor as good or bad, according to the observer's social norm.
If the social norm is L6, each observer $o$ judges the donor as follows:
\begin{equation}
\sigma_{od}' = 
\sigma_{or} \cdot \sigma_{dr},
\label{eq:L6}
\end{equation}
where the prime on the left-hand side means that it is an updated value after observing the interaction.
It is instructive to review the transition structure of the cluster dynamics induced by L6 [Fig.~\ref{fig:transition}(a)]. Consider a complete graph with $N$ vertices. Each vertex means an individual, and the set of all $N$ vertices is called $V \equiv \{v_1, \ldots, v_N\}$. 
According to the structure theorem~\cite{harary1953notion}, a balanced configuration consists of two clusters, where the clusters are defined as two maximal cliques with respect to positive links. The two clusters in a balanced configuration can be represented as $\left\{\left\{ \ldots \right\}, \left\{ \ldots \right\}\right\}$.
One of the two clusters may be empty, in which case the configuration is called paradise and represented by $\left\{ V \right\}$. If an assessment error occurs, L6 allows one of the vertices, say $v$, to separate from the others, reaching another balanced configuration represented as $\left\{\left\{v\right\}, V-\left\{v\right\}\right\}$. If we introduce another assessment error, the system can go back to paradise, or another vertex $v'$ can migrate to the cluster of $v$, forming a cluster of $\left\{ v,v' \right\}$. The important point is that the transition probabilities, represented by the arrows in Fig.~\ref{fig:transition}(a), are equal in forward and backward directions as long as every individual's self-assessment is kept good, which is satisfied within $O(1)$ Monte Carlo steps~\cite{oishi2021balanced}. This means that every balanced configuration should be observed with the same probability in equilibrium, just as every microstate is equally probable in coin tosses. Therefore, just as we see almost equal numbers of heads and tails in coin tosses, we find two clusters of almost equal sizes under the action of L6.
For this reason, we can say that L6-induced segregation is an entropic effect~\cite{bae2024exact}.

We can think of L4 as a variation of L6, in the sense that a good person should still be regarded as good even after helping a bad one according to L4. This norm is thus described as follows:
\begin{equation}
\sigma_{od}' = 
\left\{
\begin{array}{ll}
1 & \text{if $\sigma_{od} = \sigma_{dr}=-\sigma_{or}=1$}\\
\sigma_{or} \cdot \sigma_{dr} & \text{otherwise}.
\end{array}
\right.
\label{eq:L4}
\end{equation}
Adding the exception in the first line for $\sigma_{od} = \sigma_{dr}=-\sigma_{or}=1$ leads to paradise where everyone looks good to each other. The transition structure is depicted in Fig.~\ref{fig:transition}(b). Compared with L6, this norm has one-directional transitions to paradise. Even the transition between paradise and $\left\{\left\{v\right\}, V-\left\{v\right\}\right\}$ becomes effectively one-directional as $N$ increases, so we almost always observe paradise if $N \gtrsim O(10)$~\cite{bae2024exact}.
This dramatic difference implies that the degree of segregation can be controlled by determining how to judge a good person helping a bad one.

In this work, we compare two ways of interpolation between L6 and L4~\cite{masuko2017effect}. The first is to consider a uniform population, in which everyone uses a mixture of the two social norms L4 and L6 as follows:
\begin{equation}
\sigma_{od}' = 
\left\{
\begin{array}{ll}
1 & \text{with probability $p$ if $\sigma_{od} = \sigma_{dr}=-\sigma_{or}=1$}\\
\sigma_{or} \cdot \sigma_{dr} & \text{otherwise},
\end{array}
\right.
\label{eq:monomorphic}
\end{equation}
where $p=0$ and $1$ correspond to L6 and L4, respectively.
The second way is heterogeneity, according to which $k$ individuals in the population use L4 while the rest of $N-k$ individuals use L6. It is convenient to define $f \equiv k/N$.
If all individuals in the population are indexed by integers from $1$ to $N$, we may choose the first $k$ individuals as L4 players without loss of generality.
The corresponding rule is thus described as follows:
\begin{equation}
\sigma_{od}' = 
\left\{
\begin{array}{ll}
1 & \text{if $o\le k$ and $\sigma_{od} = \sigma_{dr}=-\sigma_{or}=1$}\\
\sigma_{or} \cdot \sigma_{dr} & \text{otherwise},
\end{array}
\right.
\label{eq:polymorphic}
\end{equation}
where, with slight abuse of the notation, $o$ indicates the index of the observer under consideration.

This work is organized as follows. The next section details our numerical methods and observables. We will consider the uniform and heterogeneous situations separately, as explained above. Section~\ref{sec:results} presents our Monte Carlo results, including the difference in the sizes of the two clusters, as well as the composition of each cluster in the case of heterogeneity. We discuss the implications of our findings and conclude this work in Section~\ref{sec:discuss}.

\section{Methods}

In each round of the donation game, we randomly sample two individuals, one for a donor $d$ and the other for a recipient $r$. All $\sigma_{id}$s are updated according to either Eq.~\eqref{eq:monomorphic} or Eq.~\eqref{eq:polymorphic}, depending on whether we consider a uniform population or a heterogeneous one ($i=1, \ldots, N$).
The system becomes stationary when structural balance is achieved~\cite{bae2024exact}. Every time a balanced configuration is reached, we introduce an assessment error by randomly flipping one of $\sigma_{ij}$s to allow the system to explore different balanced configurations. The number of assessment errors thus defines time steps when we consider the coarse-grained dynamics between balanced configurations.

To determine whether a given configuration is balanced, we use the following three criteria~\cite{oishi2021balanced}. Firstly, all self-assessments must be good, that is, $\sigma_{ii} = +1$ for every $i = 1,\ldots, N$. Secondly, all opinions must be symmetric, that is, $\sigma_{ij} = \sigma_{ji}$ for every pair of $i$ and $j$. If these two criteria are satisfied, we check whether the whole system is covered by two clusters. To identify one of the two clusters, we find the unique maximal clique to which the vertex with index $i=1$ belongs, by using the Bron-Kerbosch algorithm implemented in the \pyth{networkx} package~\cite{bron1973algorithm}. This cluster will be called primary henceforth. We then find an enemy of this focal vertex and restart the algorithm there to identify the other cluster, which we will call secondary.
The size of a cluster is defined as the number of its member vertices. Let $s_1$ and $s_2$ denote the sizes of the primary and secondary clusters, respectively.
Note that our algorithm outputs $s_1=N$ and $s_2=0$ in paradise, which means that $s_1 \in [1,N]$ whereas $s_2 \in [0,N-1]$, but this asymmetry should become negligible for $N \gg 1$.
If we look at the magnitude of their difference, $\eta \equiv \lvert s_1 - s_2 \rvert$, it is straightforward to see that $m \equiv \eta/N$ approaches one under L4, whereas it should be on the order of $N^{-1/2}$ under L6. As usual, we also compute Binder's cumulant defined by
\begin{equation}
U \equiv 1 - \frac{\langle m^4 \rangle}{3\langle m^2 \rangle^2},
\label{eq:binder}
\end{equation}
where the angle brackets mean time average, where the unit of time is defined by the interval for a system to move between balanced states.

If we deal with the heterogeneous situation, we can also check how the L4 and L6 players are distributed in the system. In general, L4 players will be divided into $k_1$ and $k_2$ between the two clusters, where the total number $k = k_1+k_2$ is fixed. For example, one could guess that L6 players are almost equally divided between the two clusters, while L4 players exist in only one of them, making $s_1 \approx (N+k)/2$ and $s_2 \approx (N-k)/2$. In the next section, we will check whether this guess is correct.

\section{Results}
\label{sec:results}

\begin{figure}
\includegraphics[width=0.45\columnwidth]{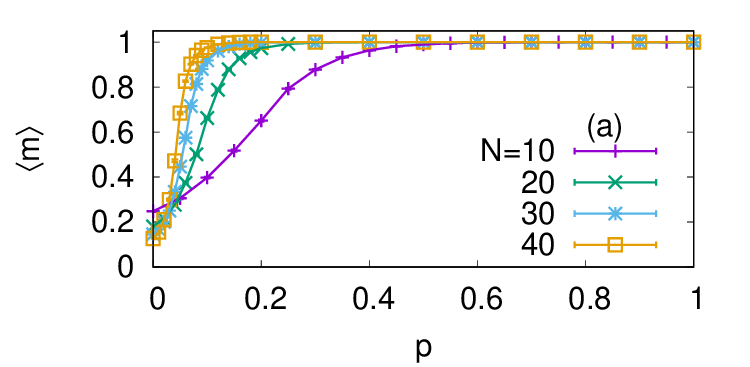}
\includegraphics[width=0.45\columnwidth]{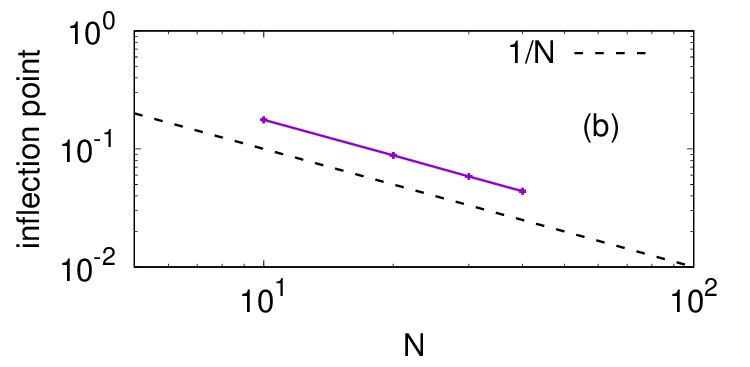}
\caption{(a) $\langle m \rangle$ as a function of $p$ at each different value of $N$ in a uniform population. We have measured the time average of $m$ over $T$ time steps, where $T=3\times 10^2$ for $N=10$ and $T=2\times 10^3$ for $N=40$, and repeated this measurement $S=10^3$ times for each data point. The error bars are smaller than the symbol size. (b) Inflection points of the curves in (a), obtained by fitting the hyperbolic tangent function to the data. For comparison, we have plotted $1/N$ as the dotted line.}
\label{fig:prob}
\end{figure}

\begin{figure}
\includegraphics[width=0.45\columnwidth]{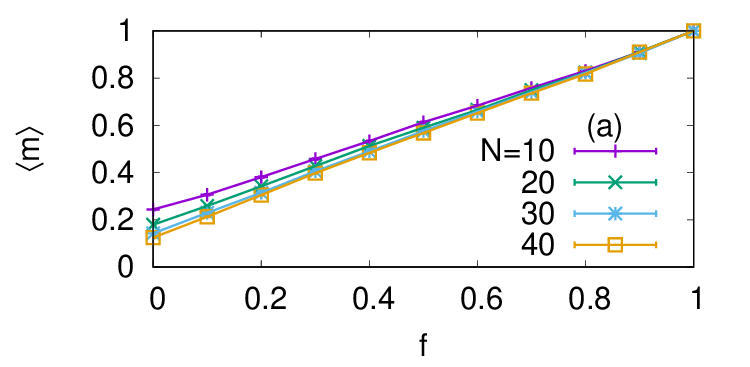}
\includegraphics[width=0.45\columnwidth]{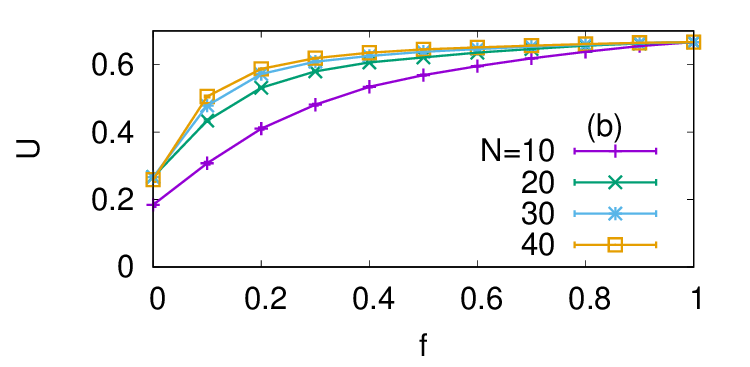}
\caption{(a) $\langle m \rangle$ as a function of $f \equiv k/N$ at each different value of $N$ in a heterogeneous population. The simulation parameters are the same as in Fig.~\ref{fig:prob}. (b) Binder's cumulant as defined in Eq.~\eqref{eq:binder}.}
\label{fig:quench}
\end{figure}

In a uniform population, even a small probability of using L4 can exert a substantial impact at the macroscopic level. As shown in Fig.~\ref{fig:prob}, our numerical data indicate that we will end up in paradise as long as $Np$ is sufficiently greater than $O(1)$. This result is qualitatively understandable because the transition structure of this probabilistic mixture of L6 and L4 must be similar to that in Fig.~\ref{fig:transition}(b), including the one-directional arrows, although the transition probabilities are different.
In contrast, in the heterogeneous case, $\langle m \rangle$ increases almost linearly with the number of L4 players, $k$ [Fig.~\ref{fig:quench}(a)]. Although $\langle m \rangle$ tends to decrease with $N$, the speed of decrease slows down and Binder's cumulant $U$ increases with $N$ [Fig.~\ref{fig:quench}(b)], indicating that $\langle m \rangle$ will remain finite for nonzero values of $f \equiv k/N$ even in the large-$N$ limit.

\begin{figure}
\includegraphics[width=0.45\columnwidth]{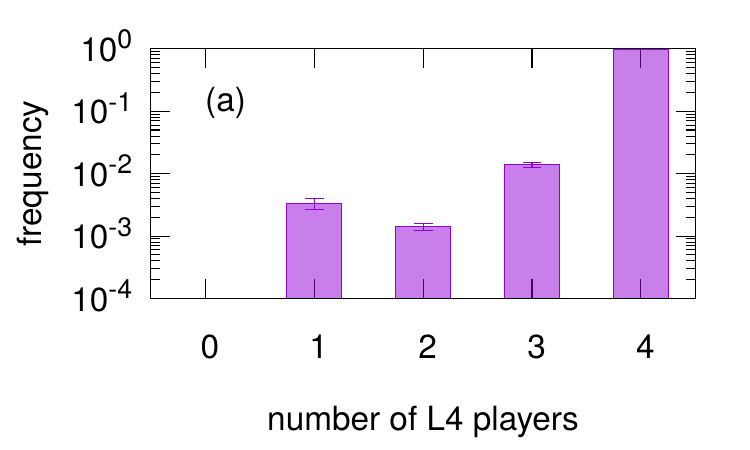}
\includegraphics[width=0.45\columnwidth]{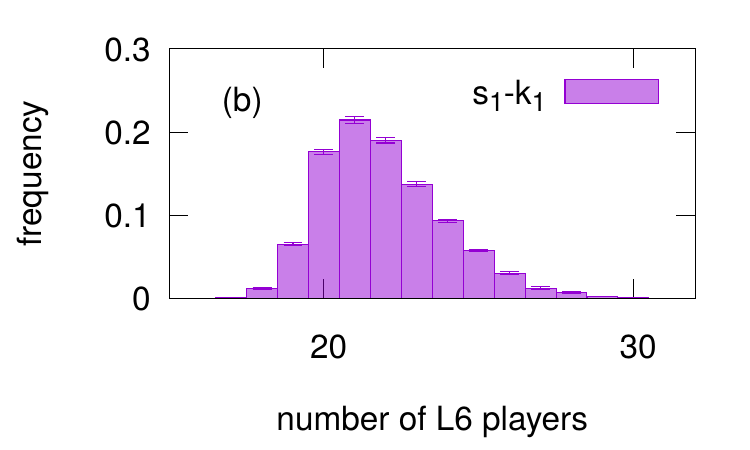}
\includegraphics[width=0.45\columnwidth]{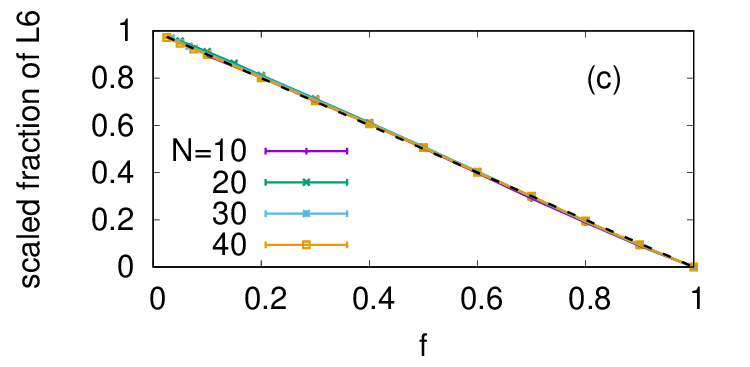}
\includegraphics[width=0.45\columnwidth]{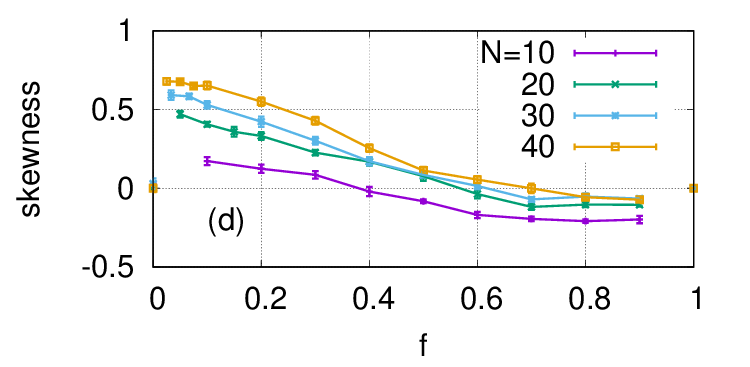}
\caption{Composition of the clusters. (a) Frequency distribution of the number of L4 players in the primary cluster, when $k=4$ and $N=10$. Note the logarithmic scale on the vertical axis. (b) Frequency distribution of the number of L6 players in the primary cluster, when $k=1$ and $N=40$.
(c) Scaled mean fraction of L6 players belonging to the primary cluster [Eq.~\eqref{eq:rescale}]. The dotted line shows $1-f = 1-k/N$.
(d) Skewness in the frequency distribution of $s_1-k_1$ as shown in (b) for various values of $N$ and $f$. The skewness is trivially zero at $f=0$ and $1$.
For each panel, we have calculated the averages and the error bars from $10^4$ samples.}
\label{fig:hist}
\end{figure}

In terms of the composition of the clusters, our numerical simulations show that all L4 players are often found together in a single cluster, that is, with $k_1=k$ and $k_2=0$. This is plausible as we already know that an L4 population prefers to form a single cluster. In Fig.~\ref{fig:hist}(a), we have plotted the frequency distribution of $k_1$ in a population with $N=10$, when there are $k=4$ individuals playing L4. Even for such a small population, the probability of observing deviation from $k_1=k$ is less than 2\%, and it becomes less and less probable as the population size $N$ increases. Therefore, if we look at the composition, the most probable configuration is such that the primary cluster consists of $k$ L4 players with some L6 players, while the rest of the L6 players make up the secondary cluster.

In addition, even L6 players are not equally divided between the two clusters, as one can see from Fig.~\ref{fig:hist}(b) that the average of the number of L6 players in the primary cluster, that is, $s_1-k_1$, is systematically greater than $(N-k)/2$. According to our simulations, it is better approximated as
\begin{equation}
\langle s_1-k_1 \rangle \approx \left(\frac{1.1N}{N-1}\right) \left(\frac{N-k}{2}\right),
\label{eq:approx}
\end{equation}
which means that
\begin{equation}
\frac{2(N-1)}{1.1N^2} \langle s_1-k_1 \rangle \approx 1-f,
\label{eq:rescale}
\end{equation}
as shown in Fig.~\ref{fig:hist}(c). Note that the prefactor $1.1N/(N-1)$ on the right-hand side of Eq.~\ref{eq:approx} is always greater than one.
We furthermore point out that the fluctuation does not follow the Gaussian distribution, and even a single L6 player ($k=1$) induces this non-Gaussian fluctuation. As shown in Fig.~\ref{fig:hist}(d), the skewness as a sign of non-Gaussianity becomes more pronounced for larger populations.

\section{Discussion and Summary}
\label{sec:discuss}

L6 is a simple and powerful norm. However, if assessments are made privately, this social norm divides the society into two mutually hostile groups. L4 has been proposed as a remedy that just alters the judgment of a good person helping a bad one. Even if this proposed remedy is approved, we can reasonably guess that the adoption of L4 will be gradual, considering its cognitive complexity [Eq.~\eqref{eq:L4}] and the inherent resistance to changes in a social norm.

In this work, we have checked two scenarios of partial adoption of L4. One of them is based on a uniform setting, where everyone uses a probabilistic mixture of L6 and L4. In this case, as long as the population size is large enough, we can almost always reach paradise because the effect of L4 eventually becomes dominant. The other setting is heterogeneity, where some have adopted L4 while the rest keep using L6. If this is the case, paradise will be unreachable until almost all players adopt L4.
Therefore, if we have to overcome the social divide induced by L6, it is better for everyone to change a little toward L4 than for the majority of the population to perfectly adopt L4, leaving the others the same as before. The condition of $Np \gtrsim O(1)$ in Fig.~\ref{fig:prob}(b) implies that each person only has to apply L4 to a handful of people, regardless of the total size of the population. As a result, individual behavior remains the same most of the time, but the global pattern of society changes.

In the heterogeneous setting, L4 players tend to gather in one of the clusters, and the presence of L4 players affects the dynamics of L6 players in a number of ways. Firstly, more than half of the L6 players join the primary cluster containing the L4 players [Eq.~\eqref{eq:approx}].
Secondly, even if a single L4 player exists in the cluster, it is enough to skew the frequency distribution of the number of accompanying L6 players [Fig.~\ref{fig:hist}(b) and (d)]. The latter finding suggests an intriguing possibility that information can be obtained on the composition of social norms in a group of people just by observing the skewness of its size distribution. 
Lastly, from an individual perspective, it will be beneficial to belong to the larger cluster consisting of L4 and L6 players because this gives a higher chance of receiving a donation. If this effect is taken into account, one could well be motivated to adopt L4, e.g., through an imitation process. We eventually obtain a pure L4 population in this way, but it will be a slow process compared to the uniform solution.

Generally speaking, our study demonstrates that when a norm is newly introduced and coexists with the existing one, the globally emerging pattern can be highly dependent on how it is introduced, i.e., whether it is a single step taken together or ten taken alone. Our example of L4 and L6 clearly shows the case, although the two norms differ only in terms of sternness when judging a good person who helps the bad.

\begin{acknowledgments}
This work was supported by the National Research Foundation of Korea (NRF) grant 
funded by the Korea government (MSIT) (RS-2025-23523609).
This work was supported by the Global Joint Research Program funded by Pukyong National University (202506730001).
The authors thank APCTP, Pohang, Korea, for their hospitality during the conference [APCTP-2025-C12], from which this work greatly benefited.
\end{acknowledgments}

\bibliography{apssamp}

\begin{thebibliography}{9}%
\makeatletter
\providecommand \@ifxundefined [1]{%
 \@ifx{#1\undefined}
}%
\providecommand \@ifnum [1]{%
 \ifnum #1\expandafter \@firstoftwo
 \else \expandafter \@secondoftwo
 \fi
}%
\providecommand \@ifx [1]{%
 \ifx #1\expandafter \@firstoftwo
 \else \expandafter \@secondoftwo
 \fi
}%
\providecommand \natexlab [1]{#1}%
\providecommand \enquote  [1]{``#1''}%
\providecommand \bibnamefont  [1]{#1}%
\providecommand \bibfnamefont [1]{#1}%
\providecommand \citenamefont [1]{#1}%
\providecommand \href@noop [0]{\@secondoftwo}%
\providecommand \href [0]{\begingroup \@sanitize@url \@href}%
\providecommand \@href[1]{\@@startlink{#1}\@@href}%
\providecommand \@@href[1]{\endgroup#1\@@endlink}%
\providecommand \@sanitize@url [0]{\catcode `\\12\catcode `\$12\catcode `\&12\catcode `\#12\catcode `\^12\catcode `\_12\catcode `\%12\relax}%
\providecommand \@@startlink[1]{}%
\providecommand \@@endlink[0]{}%
\providecommand \url  [0]{\begingroup\@sanitize@url \@url }%
\providecommand \@url [1]{\endgroup\@href {#1}{\urlprefix }}%
\providecommand \urlprefix  [0]{URL }%
\providecommand \Eprint [0]{\href }%
\providecommand \doibase [0]{http://dx.doi.org/}%
\providecommand \selectlanguage [0]{\@gobble}%
\providecommand \bibinfo  [0]{\@secondoftwo}%
\providecommand \bibfield  [0]{\@secondoftwo}%
\providecommand \translation [1]{[#1]}%
\providecommand \BibitemOpen [0]{}%
\providecommand \bibitemStop [0]{}%
\providecommand \bibitemNoStop [0]{.\EOS\space}%
\providecommand \EOS [0]{\spacefactor3000\relax}%
\providecommand \BibitemShut  [1]{\csname bibitem#1\endcsname}%
\let\auto@bib@innerbib\@empty
\bibitem [{\citenamefont {Oishi}\ \emph {et~al.}(2013)\citenamefont {Oishi}, \citenamefont {Shimada},\ and\ \citenamefont {Ito}}]{oishi2013group}%
  \BibitemOpen
  \bibfield  {author} {\bibinfo {author} {\bibfnamefont {K.}~\bibnamefont {Oishi}}, \bibinfo {author} {\bibfnamefont {T.}~\bibnamefont {Shimada}}, \ and\ \bibinfo {author} {\bibfnamefont {N.}~\bibnamefont {Ito}},\ }\href@noop {} {\bibfield  {journal} {\bibinfo  {journal} {Phys. Rev. E}\ }\textbf {\bibinfo {volume} {87}},\ \bibinfo {pages} {030801} (\bibinfo {year} {2013})}\BibitemShut {NoStop}%
\bibitem [{\citenamefont {Oishi}\ \emph {et~al.}(2021)\citenamefont {Oishi}, \citenamefont {Miyano}, \citenamefont {Kaski},\ and\ \citenamefont {Shimada}}]{oishi2021balanced}%
  \BibitemOpen
  \bibfield  {author} {\bibinfo {author} {\bibfnamefont {K.}~\bibnamefont {Oishi}}, \bibinfo {author} {\bibfnamefont {S.}~\bibnamefont {Miyano}}, \bibinfo {author} {\bibfnamefont {K.}~\bibnamefont {Kaski}}, \ and\ \bibinfo {author} {\bibfnamefont {T.}~\bibnamefont {Shimada}},\ }\href@noop {} {\bibfield  {journal} {\bibinfo  {journal} {Phys. Rev. E}\ }\textbf {\bibinfo {volume} {104}},\ \bibinfo {pages} {024310} (\bibinfo {year} {2021})}\BibitemShut {NoStop}%
\bibitem [{\citenamefont {Bae}\ \emph {et~al.}(2024)\citenamefont {Bae}, \citenamefont {Shimada},\ and\ \citenamefont {Baek}}]{bae2024exact}%
  \BibitemOpen
  \bibfield  {author} {\bibinfo {author} {\bibfnamefont {M.}~\bibnamefont {Bae}}, \bibinfo {author} {\bibfnamefont {T.}~\bibnamefont {Shimada}}, \ and\ \bibinfo {author} {\bibfnamefont {S.~K.}\ \bibnamefont {Baek}},\ }\href@noop {} {\bibfield  {journal} {\bibinfo  {journal} {Phys. Rev. E}\ }\textbf {\bibinfo {volume} {110}},\ \bibinfo {pages} {L052301} (\bibinfo {year} {2024})}\BibitemShut {NoStop}%
\bibitem [{\citenamefont {Heider}(1946)}]{heider1946attitudes}%
  \BibitemOpen
  \bibfield  {author} {\bibinfo {author} {\bibfnamefont {F.}~\bibnamefont {Heider}},\ }\href@noop {} {\bibfield  {journal} {\bibinfo  {journal} {J. Psychol.}\ }\textbf {\bibinfo {volume} {21}},\ \bibinfo {pages} {107} (\bibinfo {year} {1946})}\BibitemShut {NoStop}%
\bibitem [{\citenamefont {Harary}(1953)}]{harary1953notion}%
  \BibitemOpen
  \bibfield  {author} {\bibinfo {author} {\bibfnamefont {F.}~\bibnamefont {Harary}},\ }\href@noop {} {\bibfield  {journal} {\bibinfo  {journal} {Mich. Math. J.}\ }\textbf {\bibinfo {volume} {2}},\ \bibinfo {pages} {143} (\bibinfo {year} {1953})}\BibitemShut {NoStop}%
\bibitem [{\citenamefont {Cartwright}\ and\ \citenamefont {Harary}(1956)}]{cartwright1956structural}%
  \BibitemOpen
  \bibfield  {author} {\bibinfo {author} {\bibfnamefont {D.}~\bibnamefont {Cartwright}}\ and\ \bibinfo {author} {\bibfnamefont {F.}~\bibnamefont {Harary}},\ }\href@noop {} {\bibfield  {journal} {\bibinfo  {journal} {Psychol. Rev.}\ }\textbf {\bibinfo {volume} {63}},\ \bibinfo {pages} {277} (\bibinfo {year} {1956})}\BibitemShut {NoStop}%
\bibitem [{\citenamefont {Pacheco}\ \emph {et~al.}(2006)\citenamefont {Pacheco}, \citenamefont {Santos},\ and\ \citenamefont {Chalub}}]{pacheco2006stern}%
  \BibitemOpen
  \bibfield  {author} {\bibinfo {author} {\bibfnamefont {J.~M.}\ \bibnamefont {Pacheco}}, \bibinfo {author} {\bibfnamefont {F.~C.}\ \bibnamefont {Santos}}, \ and\ \bibinfo {author} {\bibfnamefont {F.~A.~C.}\ \bibnamefont {Chalub}},\ }\href@noop {} {\bibfield  {journal} {\bibinfo  {journal} {PLoS Comput. Biol.}\ }\textbf {\bibinfo {volume} {2}},\ \bibinfo {pages} {e178} (\bibinfo {year} {2006})}\BibitemShut {NoStop}%
\bibitem [{\citenamefont {Masuko}\ \emph {et~al.}(2017)\citenamefont {Masuko}, \citenamefont {Hiraoka}, \citenamefont {Ito},\ and\ \citenamefont {Shimada}}]{masuko2017effect}%
  \BibitemOpen
  \bibfield  {author} {\bibinfo {author} {\bibfnamefont {M.}~\bibnamefont {Masuko}}, \bibinfo {author} {\bibfnamefont {T.}~\bibnamefont {Hiraoka}}, \bibinfo {author} {\bibfnamefont {N.}~\bibnamefont {Ito}}, \ and\ \bibinfo {author} {\bibfnamefont {T.}~\bibnamefont {Shimada}},\ }\href@noop {} {\bibfield  {journal} {\bibinfo  {journal} {J. Phys. Soc. Jpn.}\ }\textbf {\bibinfo {volume} {86}},\ \bibinfo {pages} {085002} (\bibinfo {year} {2017})}\BibitemShut {NoStop}%
\bibitem [{\citenamefont {Bron}\ and\ \citenamefont {Kerbosch}(1973)}]{bron1973algorithm}%
  \BibitemOpen
  \bibfield  {author} {\bibinfo {author} {\bibfnamefont {C.}~\bibnamefont {Bron}}\ and\ \bibinfo {author} {\bibfnamefont {J.}~\bibnamefont {Kerbosch}},\ }\href@noop {} {\bibfield  {journal} {\bibinfo  {journal} {Commun. ACM}\ }\textbf {\bibinfo {volume} {16}},\ \bibinfo {pages} {575} (\bibinfo {year} {1973})}\BibitemShut {NoStop}%
\end{thebibliography}%

\end{document}